\pdfoutput=1
\documentclass{amsart}        
\usepackage{graphicx}
\usepackage{euscript}
\usepackage{verbatim}
\begin{document}

\title{Vital Rates from the Action of Mutation Accumulation}

\author{Kenneth W. Wachter}

\address{   Department of Demography, University of California, Berkeley, 2232 Piedmont Avenue,
              Berkeley, CA. 94720-2120, USA}              
              \email{wachter@demog.berkeley.edu}           
\author{David R. Steinsaltz}
\address{               Department of Statistics,
               University of Oxford,
               1 South Parks Road,
               Oxford, Oxfordshire, OX13TG, UK}
               \email{steinsal@stats.ox.ac.uk } 
\author{Steven N. Evans}
\address{Department of Statistics,
               University of California, Berkeley,
               367 Evans Hall,
               Berkeley, CA. 94720-3860, USA}
               \email{evans@stat.berkeley.edu } 

\thanks{This work has been supported by Grant AG-P01-008454 from
the U.S. National Institute on Aging, and the research of
the third author has been supported in part by Grant DMS-0405778
from the U.S. National Science Foundation.  }

\begin{abstract}
New models for evolutionary processes of mutation 
accumulation allow hypotheses about the age-specificity 
of mutational effects to be translated into predictions 
of heterogeneous population hazard functions.  
We apply these models to questions in the biodemography
of longevity, including proposed explanations of Gompertz hazards
and mortality plateaus.
\end{abstract}

\maketitle


%

\section{
Mutation Accumulation}
\label{sec.accumulation}

\par\indent

Why are flies, worms, and humans subject to laws
of age-specific adult mortality that are uncannily similar
in shape?   After suitable species-specific changes in
scale, organisms with different environments,
life histories, body plans, and lifespans turn out 
to resemble each other in the statistics of their demise.  
Similarities are typically expressed in terms of hazard
functions.  The hazard function is a summary measure of rates 
of death by age across a population, equal to the negative
slope of the logarithm of the population survivorship function.    
Hazard functions for populations 
from many species show two of the same features,
exponential increase with age over a stretch
of ages and attenuated increase over later ages
generating the visual appearance of a plateau.
The recognition of these commonalities goes back at least as far as \cite{PM35}; the generalisation and quantitative elaboration 
have been signal achievements of the new biodemography,
summed up by \cite{jV98,jC03, zeus, CT03}.

Explanations for shared features of
senescent mortality across species are sought in
considerations from reliability engineering,
from optimal life-history theory,  and from
evolutionary processes of antagonistic pleiotropy
and mutation accumulation. 
Reliability engineering is a {\em functional} 
approach to senescence, picturing the organism 
as a machine with some component structure, attempting 
to derive the failure modes of the whole from 
some presumably simpler failure modes of the components. 
The aim is usually to draw inferences from qualitative 
classes of structures to general shapes of mortality curves. 
The enterprise is considered successful if the broad 
features common to many real-world mortality rates 
are reproduced in the model. Some examples
are \cite{SM60,rR78,lG78,GG91,WF01,FE06}.


Functional models start 
from the structure of the organism, while
evolutionary models pose prior questions: 
What kind of machine is the organism, 
and why is it put together the way it is? 
Many functional models lead to the same general 
pattern for mortality rates, after all, 
and each generic class of models can yield diverse 
shapes of age-specific mortality. Optimal life-history
approaches try to narrow down the choices {\em a priori}, 
by explaining why a given structural framework, 
or a certain choice of parameters within the structural 
framework, might be evolutionarily preferred. 
Much work in this area (for 
example, \cite{mR85,SG88,pH99,kW99,WD03}) 
builds on the concept of {\em antagonistic pleiotropy}, 
introduced into the theory of senescence by George \cite{gW57}. 
Williams held that early reproduction and late survival
would be negatively 
associated through direct genetic mechanisms.
Recent research often abstracts from the genetic term ``pleiotropy'',
to contemplation of more general trade-offs and compromises 
that operate across time within the lifetime of an organism 
and across generations (cf. \cite{TGGS98,pH01,Campisi03,tN03}).

The other side of the conventional evolutionary 
theory of aging, called {\em mutation accumulation}, 
views senescence not as an optimal trade-off between
early- and late-life reproductive success, but rather as 
the age-specific effect of genetic load, 
a concept developed by Peter \cite{pM52}.
Ongoing random mutation spews mostly deleterious 
changes into the genome. Since the only genetic 
``repair mechanism'' is the death of the 
organism carrying the defect, there is perpetually an overhang of 
deaths not yet realized, stretching from the time of the initial
mutation until all descendants have died from the effect of the allele. 
Less nocive mutations linger.  Since an 
individual may not live long enough to experience the harm from 
a late-acting mutation, this provides another process through
which natural selection reshapes demographic schedules.
At equilibrium,  mortality rates trend upward with age 
in proportion to the weakening force of selection.   
The population observed at any given time will
be found to be genetically heterogeneous, because new 
mutations with particular age effects are scattered 
independently across the individuals in a population
and the mutations act together to alter each 
individual's internal susceptibilities to causes of death. 

All these approaches have something
to contribute to an understanding of the central
phenomenon at issue,  that risks of impairment
and death increase with age.  None of the approaches
excludes the others.
This article treats mutation accumulation,
but in a way that incorporates, as a start, 
one characteristic feature from reliability
models,  early-age concomitants of late-age
debilitation.  In future work we hope to
tackle head-on the challenge of linking 
evolutionary models with mechanistic
and physiological models.    
Trade-offs and impacts on age-specific mortality
must be embodied in complex reliability structures.
With the exception of \cite{PN00}, 
the mathematical development on both sides
has up to now lacked the flexibility 
required for a more synoptic model.




Decades of research have established,  piece by piece, a 
mathematical framework for characterizing genetic load
and the interplay between mutation, selection, and
recombination.  Developments through the end of
the Twentieth Century are  
presented in an authoritative book by \cite{rB00}.  
Early achievements addressed single-locus 
and several-locus systems with rich genetic structure, 
but did not attempt to superimpose demographic 
dimensions.  During the 1990s,  Brian \cite{bC94}
succeeded in consolidating an age-specific 
demographic treatment based on a linear
approximation.
\cite{bC01}
showed that both of the tell-tale common features of
hazard functions across species,   the exponential
Gompertzian rise and the eventual onset of plateaus,
could be predicted by the linear approximate model 
from simple, minimalist assumptions.  His ideas have
attracted wide attention.


Three obstacles have hitherto blocked the  
path to a broader application of mutation-accumulation models: 
first, the limited versions of age-specific 
genetic harm under consideration,  
second, the assumption that genetic loci
affecting different ranges of ages evolve independently,
and, third,  inattention to heterogeneity. 

The early work of W. D.  \cite{wH66} 
posited mutations that apply a single bolus of mortality 
at one fixed age, what we call a ``point-mass'' model.
B. \cite{bC01} 
tried other stylized patterns: 
mortality increments within specified windows, 
in Gaussian shapes around specified centers,  
or beyond specified ages of onset.  
He also tried coupling these age-specific
patterns with an increment independent of age;
our coupling of late-age with early age effects
follows in this spirit precedent.    

A provision that late-acting effects 
carry with them some early manifestations 
is characteristic of reliability models for
senescent mortality.  A typical example 
is the model of \cite{GG01},
which posits an underlying structure of 
independent components and identifies
death with the first component failure.
Waiting time until death may have a mean or
mode late in life but its distribution will
have a left-hand tail showing up in some
early deaths.  We appropriate this feature
for our applications of mutation accumulation.   
Building evolutionary structure
directly into reliability models remains a  
project for the future.  



On the issue of independence,  
it is an essential feature of demographically-based models 
that the evolution at distinct sites fails to be 
independent {\em even if sites act independently}; 
that is, even if the mortality increment due to two alleles 
co-occurring is merely the sum of their individual effects.
To put it simply, death comes to an individual only once, 
so that any mutation that increases mortality makes
a second mutation that also increases mortality less costly, 
as measured in lost reproductive opportunity.
Linearization, as previously employed, treats multiple
mutations as though they were evolving independently 
and so misses the critical interaction effect in the cumulative
demographic impact.

On the issue of heterogeneity,  natural selection must 
have variability on which to act.  Selection can
only balance mutation when some members of the
population carry more deleterious mutant alleles
than others.    The levels of the mean counts of 
mutant alleles at equilibrium are altered by the 
variability of counts about their means,
the variability which  drives the whole 
mutation-selection process.   The feedback from
variances to means is typically suppressed by
linearization.  

All three of these imperatives,
flexible profiles for effects,  interactions,
and heterogeneity,   call for a fully 
nonlinear model, such as the one applied
here.   The need for such a model seems
to have been appreciated already in 
Brian  
Charlesworth's (2001)  
pathbreaking paper.  In his Section 4 
he sought to incorporate nonlinear interactions 
through an iterative numerical procedure, 
making survivorship at each step in time depend on
the previous mean accumulation of mutant alleles
en route to an equilibrium.   This procedure
suppresses heterogeneity and leads to different
answers from our fully nonlinear model,
but in some circumstances it generates usable 
approximations.  It highlights the importance
of nonlinear effects.  Even in the ``point-mass''
setting, nonlinearity can produce outcomes
qualitatively different from those predicted
with the linearized approach, as shown 
in \cite{SEW05} and \cite{WES08}.  
The full model also makes it possible to prove conditions
for the existence of equilibria and Walls of Death.   




In our application here,  mutant alleles arise that 
each increase age-specific mortality rates according 
to the profile of a Gamma probability density function.
The model builds in the nonlinear demographic
interactions among the accumulating mutant alleles, and takes explicit 
account of the heterogeneity in genetic endowment among individuals.  
Investigating a range of choices of parameter values, 
we show that the features of prime demographic interest,  
Gompertzian stretches and late-age plateaus,  
can be produced within this setting.  

The Gamma profiles adopted here are reminiscent
of functional forms common in reliability models,
but no precise analogy is intended.    
Our choice was guided by the idea that 
the ``essential organs'' of \cite{GG91} 
might be replaced by a large number
of ``useful organs'', of similar internal redundancy, 
whose propensity to failure could be triggered or exacerbated
by the presence of one or more mutant alleles.
Ultimately we hope it will be feasible to 
situate reliability models explicitly within
the context of mutation accumulation.
The evolutionary unified failure theory
of our aspiration would also need to incorporate elements 
of optimal life history, as well as accounting for 
the complex hierarchy of trade-offs, from the level of
single genes and organelles up to ecosystems, 
and on timescales from the milliseconds 
of RNA transcription to the millennia of evolutionary time.

For our present, more modest, purposes the Gamma family was
chosen because it has the desired property
that late-age increments in mortality are systematically
tied to early-age increments in a fashion that varies 
smoothly with the mean age of effect.  
This specification takes us beyond the highly stylized setting 
of point-mass cases,  while retaining enough familiarity
for ready interpretation.
 
We describe the model in Section \ref{sec.themodel},
the formulas that go into demographic calculations 
in Section \ref{sec.formulas}, and the detailed specification
of ingredients and parameters in Section \ref{sec.profiles}.
We present the mortality outcomes predicted by the theory in 
Section \ref{sec.predictions}.

\section{
The mutation-selection model} \
\label{sec.themodel}

Medawar's idea of mutation accumulation as a 
cause of senescence depends upon the action 
of large numbers of mutations, each with small deleterious 
effects on survival at specific ranges of age.
Mutations which affect young ages are weeded out of the 
population quickly by natural selection, because members
who carry them contribute fewer offspring to the next
generation.  Mutations affecting older individuals, with less 
reproductive potential remaining to lose, are weeded out 
less rapidly.   While weeding progresses, new 
mutations are being introduced at random into the population.  
Mutant alleles accumulate until a balance is reached 
between the force of mutation and the force of selection.
All things being equal, the less costly mutations --- those that
produce their harm later --- will be more common at equilibrium.

Our model for mutation accumulation is an infinite-population
model in continuous time with large or infinite numbers of 
genetic loci, in the tradition of a famous paper by \cite{KM66}.
The model comes in two versions.  The version applied here
incorporates what we call ``Free Recombination'',
in which recombination is assumed to operate
on a more rapid time scale than mutation and selection.  
For mathematical details we refer the reader to \cite{WES08}
and to \cite{diploid} which shows that this version can be
regarded as a limiting case of discrete-generation
models in the limit of weak selection and mutation.
A companion version in which recombination
is assumed to be negligible is developed in \cite{SEW05}.
Our two treatments of recombination bracket
a potential continuum of more complex treatments.

For each version of the model, there are analytic 
solutions available to describe entire time 
trajectories for the population. 
In this paper, we are primarily
concerned with equilibrium states.
Equilibrium states are distributions of genotypes
which are stable in time under the
joint action of mutation and selection.
In many situations, including those treated here, we
can prove that there is a unique equilibrium state,
and that this state represents the distribution to which the
population converges over time.

The accumulating mutations under study here are germ-line
mutations maintained in the genome over long stretches of
evolutionary time.  Our framework may also
have some application to somatic mutations accumulating
within the cells of an individual individual organism during its lifecourse,
but that is not our current focus. 

The model has three ingredients which must be 
specified for each application.   
First is a set of profiles for the age-specific
action of deleterious mutant alleles.
Second is a specification of the rates at which mutant
alleles of different kinds arise, the mutation 
part of mutation-selection balance.    
Third is a function determining selective cost,  
the selection part of mutation-selection balance.  
In this section we develop the framework, with
notation and formulas in the following section.
Our choices for the ingredients,
which serve as illustrations in this section,
are spelled out in detail in Section \ref{sec.profiles}.  

\begin{figure}[!ht]
\vspace{-1.5cm}
\begin{center}
\includegraphics[width=4.5in,height=5in,
 keepaspectratio=false,angle = 90]{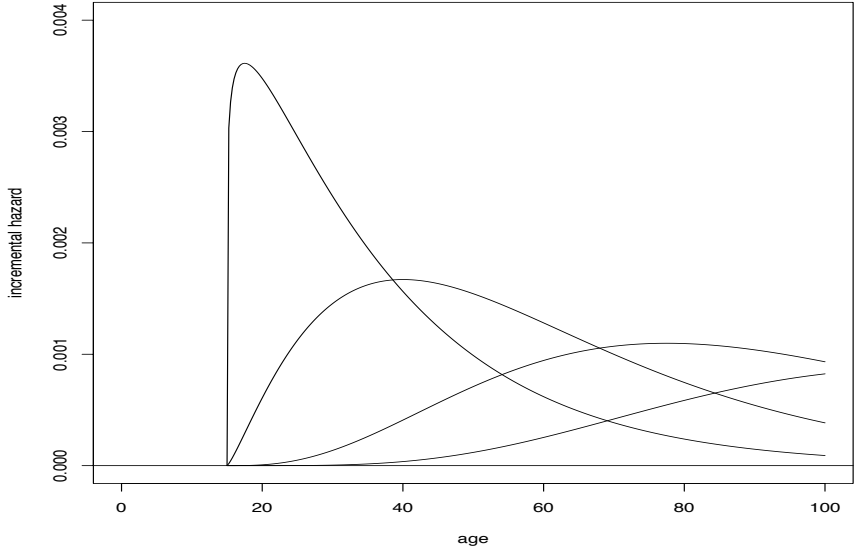}
 \vspace{-1.5cm}
\caption{\label{figgamden} 
Gamma profiles for increments to the hazard function for four
selected values of the mutation index $m$, namely $ 1.125 $, $2.250$,
$4.125$, and $ 6.000$ (from left to right as they rise from the axis).
Background parameters are $\alpha = 15 $ and $ \phi = 1/20 $,
and $ \eta = 0.100$}
\end{center}
\end{figure}

Our first ingredient is a set of profiles for 
age-specific action.   Examples of profiles
are the four functions of age in Figure \ref{figgamden},
a figure discussed further in Section ref{sec.profiles}.
A profile is added 
onto the age-specific hazard function for each 
mutant allele carried by an individual.

In general, we posit a set $\mathcal{M}$ of potential
mutations fitted with a geometric structure to allow 
us to describe the process of picking a new set 
of random mutations which are passed on to the next generation.
In our application,  the profiles form a one-parameter
family of curves,  and we can identify $\mathcal{M}$
with the interval of the real line containing the
permitted values of the parameter.  
Picking a set of random mutations comes down to
picking a random set of points from the real line,
what probability theorists call a point process,
in this case a Poisson point process.  
The important feature of the profiles is their
dependence on age.  No attempt is being made to
identify alleles with genes on chromosomes
or otherwise to model biological structures.  
Versatile age-specific structure is achieved in conjunction
with a degree of stylization in the representation of the
genome.

Our second ingredient gives rates at which mutant alleles
arise, rates expressed in general by measure $\nu$ 
on $\mathcal{M}$.   When we take $\mathcal{M}$ to be
a real interval, we can identify $\nu$ with a 
non-negative function, the density of the measure, 
and we often take $\nu$ constant over the interval
for the sake of having a neutral choice.  
Time is continuous in our model. 
It is convenient to scale the time axis so that 
one unit of time corresponds to one generation in 
discrete settings.  The rate $\nu$ is then expressed in units
of mutations per generation.
 
Our third ingredient is a selective cost function $S$.
It evaluates (on a logarithmic scale) the loss in fitness
produced by any batch of mutant alleles which an individual
may carry.   The alleles are dominant,   back-mutation 
is not allowed, and costs are evaluated under the 
assumption that age-specific fertility rates $f_x$ are
being rescaled to keep population size stationary and
are otherwise exogenous.  The general form of the model
allows fertility as well as mortality to be shaped 
endogenously by the action of deleterious mutations 
with a background level of any chosen form,  but 
those options are not pursued in this paper. 
In the present context, as discussed     
in \cite{WES08}, following \cite{bC00a}, page 930, 
there are good reasons for identifying the
selective cost of a batch of mutations with the 
resulting lifetime loss of net reproduction, 
and we do so here.  

\section{
Formulas} 
\label{sec.formulas}

We now introduce formal notation and describe how
the ingredients of our model fit together in terms of
the formulas from which predictions are calculated.  

We use the letter $g$ and the word ``genotype'' as shorthand 
to refer to the finite batch
of mutant alleles carried by a member of the population.
Alleles with the same profile of action are treated
as copies of the same allele even though they are found
at different sites in the genome.  An individual
without mutant alleles has the ``null genotype''  $ g = 0 $
with wild-type alleles at every site.

The survivorship function $\ell_x(g)$ for a
subpopulation of members with genotype $g$
is the proportion of members of the subpopulation
living beyond age $x$.   When we take the
logarithm of  $\ell_x(g)$ and multiply by $-1$,
we obtain the 
\it
cumulative hazard function,    
\rm
whose derivative,
when it exists,  is the hazard function itself, 
also called the ``force of mortality''.     
The cumulative hazard function at age $x$ is the
area under the hazard function up to age $x$.  
We work with cumulative hazard functions in order
to have simpler expressions for survivorships  $\ell_x$,
as well as to have formulas that apply without change
to discrete-age and continuous-age cases.  

We write  $\eta(m) \kappa(m,x) $ for the increment 
to the cumulative hazard function at age $x$ produced by 
allele $m$ from $\mathcal{M}$.  The cumulative profile
function $\kappa(m,x)$ --- the area up to age $x$ under 
a curve like the curves in Figure \ref{figgamden} --- represents
the shape of the age profile of mortality effects, normalised
to have total effect 1: that is, with $\int_{0}^{\infty}\kappa(m,x)dx=1$.   
The factor $\eta(m)$ then adjusts the overall size of the effect.  

To write the survivorship function, we start with an 
exogenous  baseline cumulative hazard $ \Lambda(x)$
and add to it a term $\eta(m) \kappa(m,x)$ for
each $m$ in the batch $g$ to obtain the cumulative
hazard.  We multiply the cumulative hazard by $-1$ and
apply the exponential function to obtain the survivorship.

We write capital $G$ for the random batch of
mutant alleles carried by an
individual selected at random from the population.
The count of alleles in $G$ with values of $m$ in
any given subset of $\mathcal{M}$ is thus a random variable.
An example would be the count of alleles with 
parameters between $2$ and $3$.
The mean of this random variable 
is just the population average number of mutant
alleles in the interval $(2, 3)$, and it equals the
area within the interval $(2,3)$ under a curve $\rho$ called
the {\em intensity} of $G$.

We now touch on some probability theory,  the material which 
enables us to go beyond linear approximations and treat 
interactions and heterogeneity.
The intensity $\rho $ gives information about overall genetic load,
but on its own it may not provide a complete
description of the genetics of the population.  
The function $\rho$ only specifies for each region
of $\mathcal{M}$ the population average number of
mutant alleles in that region and {\em a priori} does not
enable one to compute the proportion of the population
that have more than some number of mutant alleles
in a given region of $\mathcal{M}$ or to determine
whether a randomly chosen individual who happens
to have larger than expected numbers of mutant alleles
in one region is more or less likely to have a larger than
expected number in another region.
In this model, there is a distribution of genotypes,
which are batches of mutant alleles from
$\mathcal{M}$; the intensity only describes the overall
frequency of each mutation, with no information
about its genetic partners.

When selective costs are linear --- effectively,
the {\em non-epistatic} case in which
distinct loci evolve independently --- the genotype
distribution is a {\em Poisson random measure},
a mathematical construct whose properties are described,
for instance, in \cite{oK83}.
Intuitively, the genotype of an individual
sampled at random from the population
can be described by going through $\mathcal{M}$
point by point, and taking mutations independently at random
with probabilities governed by $\rho$.
In the setting of interest to us, however, when
the selective cost is nonlinear, there will
be a complex structure of interactions between mutations.

It is surprising, then, that the simple Poisson structure
returns, regardless of the complexity of the epistasis,
in our model with Free Recombination, as shown in \cite{diploid}.
While distinct loci now do not evolve independently,
the distribution at any given time {\em does} have a
the structure of a Poisson random measure and is completely described
by the intensity alone.    The population average or ``expectation
value'' of any quantity of interest depends only on $\rho$; 
we use the notation   $\mathbb{E}_{\rho}$.

Our goal, then, is to determine the intensity $\rho(m)$ of mutations
at equilibrium.  From it we can find the expected (aggregate)
population survival curve   $\mathbb{E}_{\rho} \, [ \ell_x(G)]$,
the proportion of the whole population living
beyond age $x$.  

The selective cost $S(g)$ for genotype $g$ is calculated under our assumption
of zero population growth and is given by the difference in
its Net Reproduction Ratio from the Net Reproduction Ratio for
the null genotype:   
\begin{equation}
\label{E:Sofg}
S(g) :=   \int \, f_x \, \ell_x(0) \, dx - \int \, f_x \, \ell_x(g) \, dx.
\end{equation} 
Survivorship for genotype $g$ is given by 
\begin{equation}
\label{E:ellxofg}
\ell_x(g)  =   \exp \left(- \Lambda(x) -  \sum_{m \in g} \, \rho(m) 
                 \eta(m) \kappa(m,x)
                    \right) .
\end{equation}
Section 3 in \cite{WES08}
shows that the general formulas in \cite{diploid} along with properties
of Poisson point processes imply that aggregate survivorship is 
given by 
\begin{equation} \label{E:Elxg}
\mathbb{E}_{\rho} \, [  \ell_x(G)]    =  \ell_x(0) \exp \left(
    - \int ( 1- e^{-\eta \,  \kappa(m, x)}) \,  \rho(m) \, dm  \right ).
\end{equation}

The slope of minus the logarithm of the left-hand side is
the population hazard.  The increment to the cumulative population hazard
due to the accumulation of copies of allele $m$ can be written
\begin{equation} \label{E:Hmx}
H(m,x) =  ( 1- e^{-\eta \,  \kappa(m, x)}) \,  \rho(m).
\end{equation}
In this way the left-hand side of \eqref{E:Elxg} takes
the form $ \ell_x(0) \exp( - \int H(m,x) dm ) $.

The equilibrium intensity $\rho$ has to satisfy
\begin{equation}
\label{E:kapintegral}
0  =   \nu(m)  - \rho(m) \, \int \, ( 1 - e^{-\eta \kappa(m,x)} )
                       \, f_x \mathbb{E}_{\rho} \,[ \ell_x(G) ] \, dx
\end{equation}

The equation \eqref{E:kapintegral}
can be solved numerically by an iterative scheme.
We start out with $\rho_0 \equiv 0$,
corresponding to the null genotype and, supposing we have
already constructed the approximate
solutions $\rho^0, \ldots, \rho^n$, define $\rho^{n+1}$ by
\begin{equation}
\rho^{n+1}(m) := \nu(m) \bigg / \left[\int \, ( 1 - e^{-\eta \kappa(m,x)} )
             \, f_x \mathbb{E}_{\rho^n} \,[ \ell_x(G) ] \, dx \right].
\end{equation}
Under appropriate conditions, it is possible to prove that
this sequence does in fact converge to a solution of
\eqref{E:kapintegral}.
For our numerical calculations,  we approximate the
continuous range of values of $m$ by a grid with one thousand
points and evaluate integrals over age by a grid with steps of $0.10$ years.
Calculations are implemented in the open-source R Statistical System
based on the computer system S developed at Bell Laboratories.

\section{
Specifications} 
\label{sec.profiles}

We now turn to the detailed specification of the 
cases treated in this paper,   going into the
particular choices for the three ingredients
of the model and accompanying parameters.

The first ingredient, the profiles for mutational 
action,  have been introduced in Section \ref{sec.accumulation}.   
We set 
\begin{equation}\label{E:Gamma}
\kappa(m,x) = 
\begin{cases} 
 ( 1 / \Gamma(m) )\,  \int_{\alpha}^x \, \phi^{m} \, 
    (y-\alpha)^{m-1} \, e^{-\phi (y-\alpha)} \, dy,& \quad  x \ge \alpha, \\
     0,& \quad x < \alpha.
\end{cases}
\end{equation} 
The profile function $\kappa(m,x) $ is 
the {\em cumulative} distribution function for a shifted Gamma 
probability distribution.   The Gamma shape
parameter equals the index value $m$ and varies
from allele to allele.  The Gamma rate parameter $\phi$
is the same for all alleles.  The shift $\alpha$ for
the origin is the age of maturity; alleles
affect only adult mortality.   The quantity $\Gamma(m)$ is
the ordinary gamma function.  Each effect is assigned
an effect size $\eta(m)$ which adjusts the strength
of the action.   

In addition to their association with reliability models,
Gamma distribution functions offer advantages of familiarity 
and flexibility.   They offer a clear
contrast to the point-mass profiles going back to
W. D. Hamilton already studied in \cite{WES08}.  
In the point-mass setting,  $\kappa(m,x)$ is 
a unit step-function and $m$ indexes the age 
at the step.   In our present setting,  higher
values of $m$ still correspond to later-acting
alleles,  but effects are spread across ages,
with wider spread for later-acting alleles.
Even late-acting alleles have some small effect
at young adult ages,  a salient difference from
the point-mass case.  The more the mutational effect 
is spread over older ages,  the lower is the selective cost, 
and the more copies there will be of $m$ on average 
when natural selection manages to balance recurrent mutation.

The mean age of action for allele $m$ is the mean 
of the shifted Gamma distribution $\alpha + m/\phi$,
the mode is  $\alpha+(m-1)/\phi$, and 
the standard deviation in age of action is $\sqrt{m}/\phi$. 


Figure \ref{figgamden} shows the shapes of the age-specific 
increments to the hazard function for four typical alleles in
our setting,  with $\eta \equiv 0.100$, $\alpha = 15 $, 
and $ m$ equal to $1.125$, $2.250$, $4.125$, and $ 6.000$. 
The value $\eta = 0.100$ is a typical standard value.  We
discuss the impact of other values in Section \ref{sec.predictions}.

Our second ingredient,  the mutation rate $\nu(m)$,  is taken
to be constant over an interval $[1, \xi] $ and zero outside
it.    In the choice of a constant mutation rate, we follow
the practice of \cite{bC01},  seeking to keep our assumption
about mutation as neutral as possible.  The total rate 
per generation $\nu_{tot} $ of the deleterious mutations
treated in the model amounts to the length of the 
interval, $ \xi - 1 $,  times the $\nu$ value in the interval.
We use $\nu_{tot}$ as a label.  It is the mean number of new
mutations per zygote per generation.  We consider cases 
with  $ \xi $ between $5$ and $7$ and $\nu_{tot}$
between $0.120$ and $ 0.170 $ per generation.   

The final ingredient of model specification is the selective cost
of a batch of mutations. We assume that mutations 
affect an individual's fitness only through
their effect on mortality rates, and that the 
cumulative mortality effects of multiple mutations are
additive contributions to the hazard function. 
Our selective cost function is a difference in 
Net Reproduction Ratios,  quantities which depend on
fertility as well as survival.  
We assume a fixed fertility schedule  $f_x $ equal 
to $0$ below an age of maturity $\alpha $ and above
a latest age at reproduction $\beta$ and equal to
a non-zero constant between these ages. 
The value of the constant is tuned to 
produce an overall stationary population size. 
For the predictions of Section \ref{sec.predictions},
$\alpha = 15 $ and $\beta = 50$. 

The inclusion of an upper age limit on fertility 
is important to the interpretation of our results.
The values of the age-specific profile $\kappa(m,x) $
for $x$ above $\beta$ are irrelevant to the selective
cost imposed by $m$ and therefore to the equilibrium
frequency of $m$,  but they make significant 
contributions to the predicted post-reproductive hazard.  
It is the association between early-age
and late-age hazards built into our family of profiles 
that drives the predicted outcomes.  Biologically,
we are assuming a correlation between young and old
ages in phenotypic effects.  Reasons for doing so,
in relation to reliability models for aging, have
been discussed in Section \ref{sec.accumulation}.  
The correlation across ages prevents the occurrence
of a Wall of Death at the end of reproduction 
and shapes the old-age hazards.  

The selective cost function also depends on the choice
of baseline survival schedule, the schedule for the null genotype. 
Following the lead of \cite{bC01},  
we assume a constant baseline hazard $\lambda$ above 
the age of maturity $\alpha$,  corresponding to a cumulative
baseline hazard $\Lambda (x) = \lambda ( x - \alpha)  $  
above $\alpha$ and zero below.  Since we are rescaling
fertility to achieve stationarity, pre-reproductive 
mortality can be ignored.  The baseline hazard  
can be taken to represent a minimum realizable rate, 
sometimes identified with the 
so-called the {\em extrinsic mortality rate}  
despite the problems inherent in this  notion
discussed by \cite{WD03}.  
As with fertility, our choice of baseline hazard is
intended to be as neutral as possible, in order to
concentrate on structure arising from the dynamics
of mutation and selection.

\section{Predictions}
\label{sec.predictions}

\begin{figure}[!ht]
\vspace{-1.5cm}
\begin{center}
\includegraphics[width=4.5in,height=4in, angle = 90]{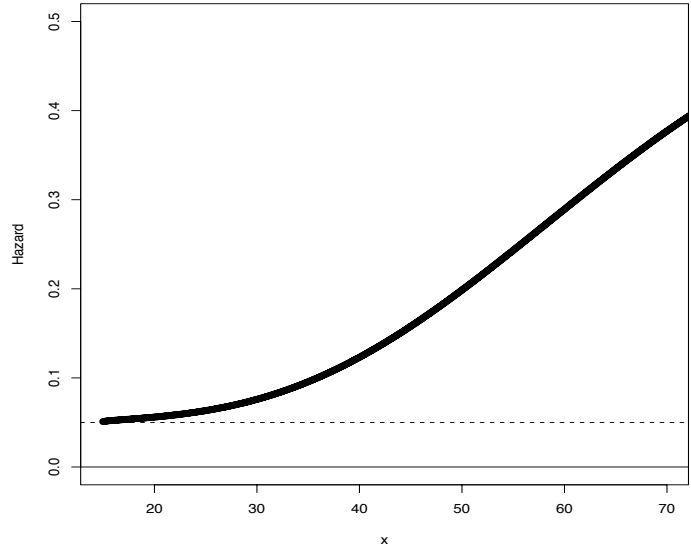}
\vspace{-1.5cm}
\caption{\label{fighaz} 
Predicted hazard for a standard example with $ \lambda = \phi = 1/20$,
$\nu_{tot} = 0.150 $, $ \xi = 6 $, $\alpha = 15$, and $\beta = 50$}
\end{center}
\end{figure}

We now examine predicted hazard functions at mutation-selection
equilibrium when the age-specific action of mutant alleles takes
the form of Gamma profiles described in Section \ref{sec.profiles}.
We begin with a case chosen to serve as a standard example,
to which we shall compare other cases.  It is illustrated in
Figure \ref{fighaz}.

For our standard example, we set  
the upper cutoff on shape parameters $\xi = 6$,
and the total mutation rate $\nu_{tot} = 0.150 $,
along with an effect size $\eta$ constant at $0.100$. 
a baseline mortality level $\lambda = 1/20$,
and a rate parameter $ \phi = 1/20$.  
The maximum increment at any one age associated with our Gamma 
profiles is then a little more than three per thousand per year. 
For the sake of analogy with human life history, 
we set the age of initial reproduction (also the age
of earliest action of the mortality profiles) to be $\alpha=15$,
and the age at end of reproduction to be $ \beta = 50$. 


Figure \ref{fighaz} shows the population
hazard rate  calculated from \eqref{E:Elxg}.  
It rises slowly from the background level and
then accelerates, giving the impression of a 
Gompertz-Makeham curve in the middle of the
age range, and straightening out at older ages.
About one in ten-thousand individuals survive
beyond age $70$. 

In this illustration, the equilibrium density of 
mutations $\rho$ turns out to be closely approximated by
an exponential function of the shape parameter,
namely $ \rho(m) \approx 0.170 \exp( 1.377 \, m) $.
On average, individuals in the population carry 
about two mutant alleles with $m < 2.0 $, a bit over a dozen 
with $ 3.5 < m < 4.0 $, and nearly three-hundred 
with $ 5.5 < m < 6.0 $, for an average total of $526$. 
The Poisson standard deviation of the total number of
alleles across individuals is around $23$.


%

The effect for a given $m$ peaks at age $ 15 + (m-1)/\phi $.  
The effects for $m < 2.0 $ are peaking before age $35$,
an age to which most members of the population survive.
The cost in net reproduction from an additional mutation
affecting these ages is high,  and selection keeps their
equilibrium representation low.   Effects for $ m > 5.5 $  
only become substantial at ages at which most individuals 
have already died. Selective costs are low and copies
persist long enough to be found at high numbers in
the population despite the rarity of new mutations. 

\begin{figure}[!h]
\vspace{-1.5cm}
\begin{center}
\includegraphics[width=4.5in,height=4in, angle = 90]{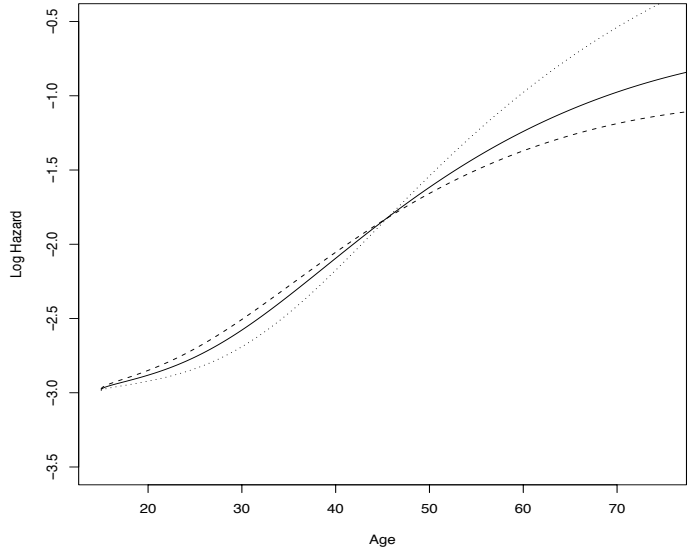}
\vspace{-1.5cm}
\caption{\label{fig3loghaz} Logarithm of predicted hazard for three cases showing early upward
bend, straight middle Gompertzian stretch,  and late downward bend.
The cases all have $\lambda = \phi = 1/20 $ and $\eta \equiv 0.100$.
The solid curve has $\nu_{tot} = 0.150$, and $\xi = 6.0$;
the dashed curve has $\nu_{tot} = 0.170$, and $\xi = 5.5$;
the dotted curve has $\nu_{tot} = 0.120$, and $\xi = 7.0$.}
\end{center}
\end{figure}

Figure \ref{fig3loghaz} shows the logarithm of
the population hazard rate for three comparative cases.
The standard example is the solid curve.
The dotted curve has $ \nu_{tot} = 0.170$ 
and an upper shape parameter cutoff of $\xi = 5.5$.
The dashed curve has $ \nu_{tot} = 0.120$ and $\xi = 7$.
These alternatives have been chosen from among cases
for which the predicted equlibrium hazard is between $0.300$ and $0.550$
at the age to which one in ten thousand survive,
respectively equal to $69.9$, $71.9$, and $66.7$ years. 

The higher hazards at old ages in the dotted curve are due
to the presence of later-acting alleles with $m$ ranging
up to $7$.  These alleles have effects whose age-specific
profiles increase throughout the range of ages to which
population members survive. The mode for $ m = 7 $ is
not reached until $135$ years.  Although the mutation 
rate is lower for the dotted curve, the shapes of the age-specific
effects lead to higher hazards toward the end of life.


We see that mutation accumulation with the given 
profiles and parameters produces a long middle 
stretch of nearly loglinear hazards,  corresponding to a 
Gompertz form. At young ages the curves are convex on
the logarithmic scale, bending upward,
as effects of mutant alleles come into play.
At older ages, the curves turn concave.
Accumulation of mutational effects concentrated at 
late ages is held in check by their small 
accompanying effects at young ages in this
specification.    

\begin{figure}[!ht]
\vspace{-2cm}
\begin{center}
\includegraphics[width=4.5in,height=5in,  angle = 90]{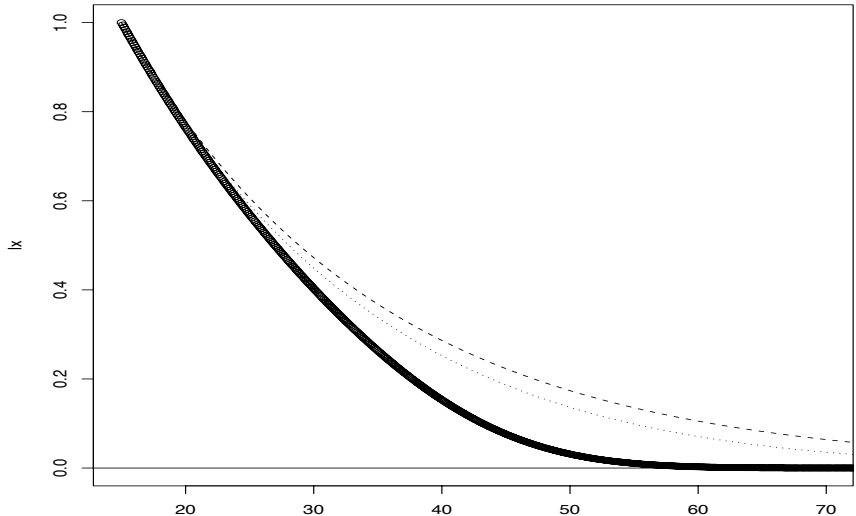}
\vspace{-2cm}
\caption{\label{figlx} 
Probability of survival by age for the 
standard example as predicted by the full nonlinear model (circles), 
by a linear approximate model (dots), and by the baseline
model}
\end{center}
\end{figure}

The interactions among effects at different ages 
taken into account by the nonlinear model turn
out to have a substantial impact on predictions,
as expected from results in \cite{WES08}.
We compare predictions
from the full nonlinear model to predictions from a
linear approximate model of the kind on which earlier
studies have relied.   Figure \ref{figlx} 
shows the population survivorship function for our
standard example in a thick line,  along with 
baseline survivorship in a dashed line,  and 
survivorship from the linear approximate model 
between them in a dotted line.   
Only about a third of the reduction in life expectancy 
from $35.0$ years to $28.8$ years due to mutation
accumulation is captured by the linear approximate model.  

\begin{figure}[!ht]
\vspace{-2cm}\includegraphics[width=4.5in,height=5in,angle=90]{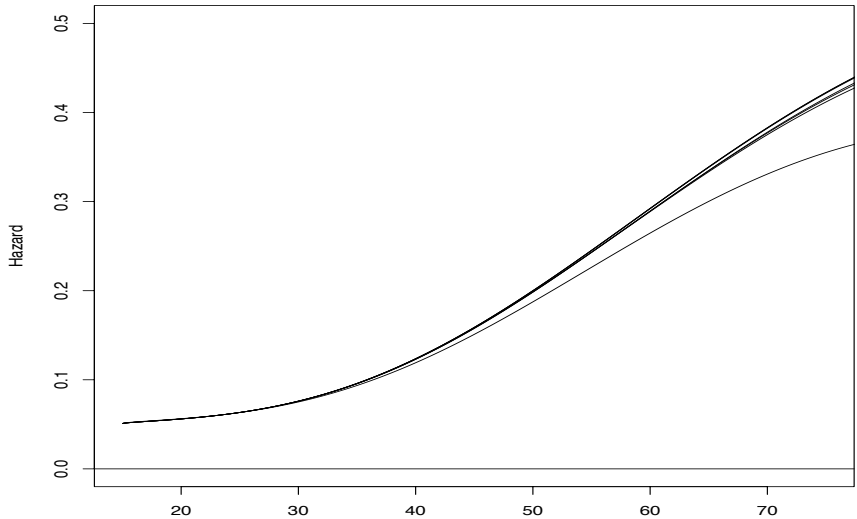}
\vspace{-2cm}
\begin{center}
\caption{\label{figeta} 
The influence of effect size is shown with predicted hazard
functions from seven cases described in the text, 
some with indistinguishable outcomes,  sharing all parameter 
values except effect size with the example of Figure \ref{figgamden}.}
\end{center}
\end{figure}

The sizes of effects,  in contrast to their shapes,
turn out to have only modest influence on the predictions.  
Alleles with smaller effects accumulate at equilibrium
in greater numbers.  Changes in the intensity $\rho$ roughly
balance changes in effect size $\eta$. Figure \ref{figeta}
shows the predicted hazard functions with parameters taken
from our standard example but with different choices of $\eta$. 

%

Three uppermost curves, almost indistinguishable, 
have constant $\eta = 0.0001$, $ \eta = 0.001$, and $\eta = 0.010$.
The mean total number of mutant alleles runs to a bit 
over five hundred thousand in the first case, fifty-thousand in the
second, and five thousand in the third.
Three slightly lower curves, also hardly distinguishable,
include our example with $\eta$ constant at $0.100$ and two examples
with changing $\eta$, one rising linearly with the shape parameter
from $0.020$ to $0.200$ and one falling linearly from $0.200$ to $0.020$.
Mean counts of alleles are $527$, $309$ and $1493$ respectively.
Small effect sizes accompany larger mean
numbers when they occur for alleles with late action.   
The lowermost curve has $\eta = 1.000$ and a mean of only $51$ alleles.
Only as $\eta$ becomes this large, outside the range of intended
application of the model,  do we see substantially different
predicted hazard functions.

The approximate invariance of predicted hazard functions with
effect sizes is an expression of Haldane's Principle,
enunciated by \cite{jH37} and discussed in terms of our nonlinear
models in \cite{WES08}.  
In Equation \eqref{E:Hmx},   the contribution $ H(m,x) $ 
from allele $m$ is nearly linear in $ \eta(m) \rho(m) $ 
for small $\eta$, so scaling $\eta(m) $ up can be nearly 
compensated, allele by allele,  by scaling $\rho(m)$.

\begin{figure}[!h]
\vspace{-2.5cm}\begin{center}
\includegraphics[width=4.5in,height=5in, angle = 90]{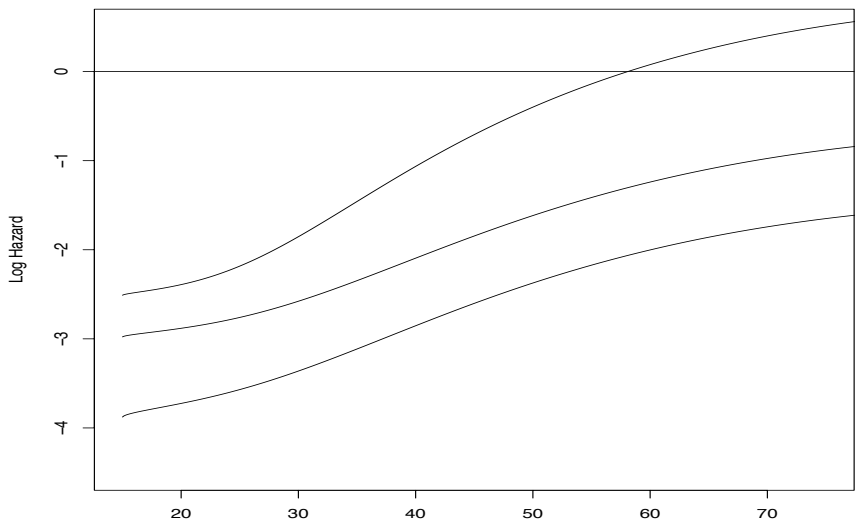}
\vspace{-2.5cm}
\caption{\label{figlam} 
The influence of the level of extrinsic hazards on the
pace of senescent mortality is shown by predicted log
hazards from three cases sharing parameters with the
standard example of Figure 1 except for $\lambda$,
which ranges, from bottom to top, over 
values  $0.020$, $0.050$, and $0.080$ }
\end{center}
\end{figure}

One of the most familiar general predictions of the
evolutionary theory of senescence is a positive relationship
between the extrinsic mortality rate
associated with unavoidable risks in natural settings such as predation and accidents
that are present even the young and healthy,
and the ``rate of senescence'' measured by the slope
of the logarithm of the hazard rate with respect to age. 
Our predictions hint at such a relationship, but only
for substantial values of the baseline hazard $\lambda$.
Figure \ref{figlam} shows the logarithms of the 
predicted equilibrium hazard for our standard set of
parameters as the level of the constant baseline
hazard is raised from $0.020$ to $0.050$ and on to $0.080$.

Slopes computed over the middle range of ages from $30$ to $50$
to which a Gompertz fit is roughly appropriate hover around $ 0.050 $
for the first two cases but rise to $ 0.074 $ as $\lambda$ increases
to $0.080$.    In cases not shown here in which non-zero
fertility extends to higher ages,  there is a closer
match between values of the slope and values of the 
parameter $\lambda$ itself, paralleling a relationship
found with linear approximate models in \cite{bC01}.

\begin{figure}[!ht]
\vspace{-1.5cm}
\begin{center}
\includegraphics[width=4.5in,height=5in, angle = 90]{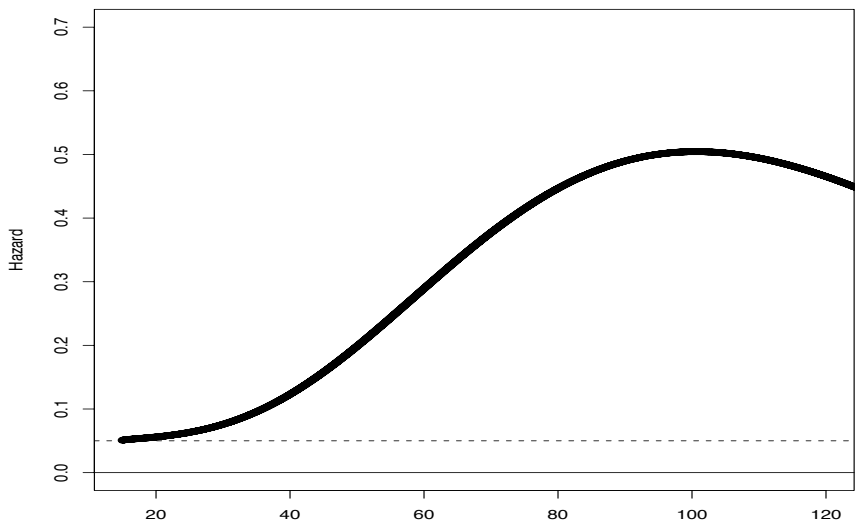} \vspace{-1.5cm}
\caption{\label{figplat} 
A plateau in the predicted hazard function at extreme
ages in the standard example of Figure \ref{fighaz} with a longer
range of ages.}
\end{center}
\end{figure}

The stretch of ages with exponentially increasing 
hazards,  corresponding to linear increase in log hazards,
visible in Figures \ref{fighaz} and \ref{fig3loghaz} 
does not extend out to extreme ages.  Attenuation of
increase is already visible in the upper ages toward right 
of those figures.  We focus on this
attenuation in Figure \ref{figplat},  which shows the
predicted hazard rate for the same standard example
of Figure \ref{fighaz} but with a horizontal axis
extending all the way out to an age of $120$ years.
The vertical scale also differs from Figure \ref{fighaz}.
Around $100$ years,  the hazard levels off, 
establishing a brief plateau phase,  and by $120$ years
a declining hazard is apparent.  Only one in ten billion
survive to $100$ years with our standard parameter 
choices,  but the plateau and subsequent decline 
are to be expected with parameters leading to milder
mortality regimes as well.

The plateau at extreme ages is due to the property of
the profiles for the age-specific action of mutant alleles
to which we have already called attention,  namely that
even alleles whose action is spread over old ages all
have some small effects at young ages.  These small
effects rein in the accumulation of late-acting mutant
alleles.  They prevent any Wall of Death; that is,
any finite age at which the hazard rate goes to infinity and survivorship
reaches zero.  Walls of Death occur in many elementary cases
for profiles with Hamilton-style, point-mass profiles,
as shown in \cite{WES08}.   The proposal for generating
plateaus by assuming some small effects at young ages
for all mutant alleles was put forward by \cite{bC01}
and shown to be valid for the linear approximate model.
We now see that these outcomes also hold in the full
nonlinear model with the particular profiles we are
studying.

In summary,  we have found that the process of mutation 
accumulation can readily produce predicted population
hazard functions with the chief features highlighted by
the cross-species comparisons of biodemographers.
It can produce a stretch of ages with an exponential, 
Gompertzian rise in hazards and it can produce 
a late-age hazard plateau.
These outcomes arise from a set of assumptions about
the age-specific action of mutant alleles that are
suggested by examples from reliability theory and
from the functional approach to the study of
senescence.  It remains, however, to develop 
comprehensive models in which the generic mutation 
accumulation machinery is driven by plausible genetic 
and physiological mechanisms, and in which age-specific 
tradeoffs are derived compellingly from reliability theory.   
It also remains to be determined
whether the examples studied here are typical, or whether they
represent peculiar outcomes of our specific choices of parameter values.
More generally, the field is open for attempts to 
characterize the conditions under which
the force of natural selection in the presence of 
recurring deleterious mutation will mold
hazard functions into familiar forms.


%

\end{document}